# High temperature superconductivity in metallic region near Mott transition

Tian De Cao

*Department of physics, Nanjing University of Information Science & Technology, Nanjing 210044, China*

The spin-singlet superconductivity without phonons is examined in consideration of correlations on an extended Hubbard model. It is shown that the superconductivity requires not only the total correlation should be strong enough but also the density of state around Fermi energy should be large enough, which shows that the high temperature superconductivity could only be found in the metallic region near the Mott metal–insulator transition (MIT). Other properties of superconductors are also discussed on these conclusions.

The superconductivity of Cu-based superconductors [1-5] occur at the region where the long-range antiferromagnetic order has been disappeared, the superconductivity of Fe-based superconductors [6-10] may also appear at the border of spin density wave (SDW) or other magnetic orders. In a word, the high temperature superconductivity usually appears in the border of the magnetic orders [11]. In addition, the superconductivity could not be appeared in good metals; examples include the heavily doped copper oxides and Au, Ag, Cu, etc. What is the factor dominating the superconductivity? Our work suggested that superconductivity may be dominated by the spin-charge correlation [12], with which various excitations could mediate the superconducting pairing, and these ideas are suggested again in following experiments, such as Park and his coauthor's work [13] which argues that magnetic and charge fluctuations coexist and produce electronic scattering that is maximal at the optimal pressure for superconductivity. However, whether superconductivity is included in the Hubbard model [14-21] has been an open question. We find that some suggestions against superconductivity are because they have introduced operators similar to the summary over lattice sites in real space; their results are in fact unreliable [22]. Thus we should find more details of superconductivities.

To consider the physics of actual materials, we extend the Hubbard model to this form

$$H = \sum_{l,l',\sigma}(t_{ll'} - \mu\delta_{ll'})d^+_{l\sigma}d_{l'\sigma} + U\sum_{l} n_{l\uparrow}n_{l\downarrow} + \frac{1}{4}\sum_{l,l',\sigma,\sigma'}V_{ll'}n_{l\sigma}n_{l'\sigma'} - \sum_{l,l'}J_{ll'}\hat{S}_{lz}\hat{S}_{l'z}, \tag{1}$$

This model includes next nearest neighbor interactions, and it can be rewritten in

$$H = \sum_{k,\sigma} \xi_k d^+_{k\sigma} d_{k\sigma} + \sum_q V(q)\hat{\rho}(q)\hat{\rho}(-q) - \sum_q J(q)\hat{S}_z(q)\hat{S}_z(-q) \qquad (2)$$

where the charge-charge interaction matrix $V(q) = U + V_0(q)$, the spin-spin interaction $J(q) = U + J_0(q)$, the charge operator $\hat{\rho}(q) = \frac{1}{2}\sum_{k,\sigma} d^+_{k+q\sigma} d_{k\sigma}$, and the spin operator $\hat{S}(q) = \frac{1}{2}\sum_{k,\sigma} \sigma d^+_{k+q\sigma} d_{k\sigma}$ in the wave vector space when we denote $k \equiv \vec{k}$. It is found that the on-site interaction $U$ contributes the model both the charge-charge and spin-spin interaction. Because $\xi_{\bar{k}} \equiv \xi_{-k} = \xi_k$, $V(\bar{q}) = V(q)$, and $J(\bar{q}) = J(q)$, we will take $f(\bar{k}) = f(k)$ for the functions depending on wave vectors.

Green's functions are defined as

$$G(k\sigma, \tau - \tau') = - <T_\tau d_{k\sigma}(\tau) d^+_{k\sigma}(\tau')>$$

$$F^+(k\sigma, \tau - \tau') = <T_\tau d^+_{k\sigma}(\tau) d^+_{\bar{k}\bar{\sigma}}(\tau')> \qquad (3)$$

$$F(k\sigma, \tau - \tau') = <T_\tau d_{\bar{k}\bar{\sigma}}(\tau) d_{k\sigma}(\tau')>$$

where the spin singlet pairing is considered. **If the effects of correlations are neglected**, we find

$$F^+(k\sigma, i\omega_n) = -\frac{\Delta(k,\sigma)}{(i\omega_n - \bar{\xi}_{k\sigma})(i\omega_n + \bar{\xi}_{k\sigma}) - \Delta^2(k,\sigma)} \qquad (4)$$

thus the BCS gap equation is

$$\Delta(k\sigma) = \sum_q [2U + J_0(k-q) + V_0(k-q)]\Delta(q\sigma) \cdot \frac{n_F(E_{q\sigma}) - n_F(-E_{q\sigma})}{2E_{q\sigma}} \qquad (5)$$

where $\tilde{\xi}_{k\sigma} = \xi_k - \frac{1}{2}\sum_q [-J(q) + V(q)]G(k+q\sigma, \tau = 0)$ and $E_{q\sigma} = \sqrt{\tilde{\xi}^2_{q\sigma} + \Delta^2(q\sigma)}$. To arrive at Eq. (5), we have taken $\bar{S}_z = 0$, and $F(k-q,\sigma,\tau=0) = F^+(k-q,\sigma,\tau=0)$ for non-ferromagnetic states. In this case, $E_{q\sigma}$ and $\Delta(k\sigma)$ do not depend on the spin index.

On the basis of Eq. (5), superconductivity requires the matrix $2U + J_0(k-q) + V_0(k-q)$ is negative at and around the Fermi surface in the wave vector space, this will require a very large antiferromagnetic exchange parameter $J$, but this condition could not be met for actual Hamiltonians. That is to say, the superconductivity associated with spin-singlet pairing on the basis of the BCS gap equation does not appear in this tight binding model.

**Considering the effects of correlations** but following the approximations presented by Abrikosov et al [23], we must establish the dynamic equations of many-particle correlation functions such as $\partial_\tau <T_\tau \hat{S}(q)d_{k+q\sigma}d^+_{k\sigma}(\tau')>$ and $\partial_\tau <T_\tau \hat{\rho}(q)d_{k+q\sigma}d^+_{k\sigma}(\tau')>$. These calculations arrive at the equations

$$[-i\omega_n + \tilde{\xi}_{k\sigma} + \sum_q \frac{P(k,q,\sigma)}{i\omega_n - \xi_{k+q}}]G(k\sigma,i\omega_n)$$

$$= -1 + \frac{V(0)<\hat{\rho}(0)>}{-i\omega_n + \xi_k} + \frac{1}{2}\sum_q \frac{\xi_{k+q} - \xi_k}{-i\omega_n + \xi_{k+q}}[J(q)+V(q)]F(k+q\sigma,\tau=0)F^+(\bar{k}\bar{\sigma},i\omega_n) \quad (6)$$

and

$$[-i\omega_n - \tilde{\xi}_{k\sigma} - \sum_q \frac{P(k,q,\sigma)}{i\omega_n + \xi_{k+q}}]F^+(k\sigma,i\omega_n)$$

$$= \frac{1}{2}\sum_q \frac{\xi_{k+q} - \xi_k}{-i\omega_n - \xi_{k+q}}[J(q)+V(q)]F^+(k-q,\sigma,\tau=0)G(\bar{k}\bar{\sigma},i\omega_n) \quad (7)$$

where

$$P(k,q,\sigma) = \frac{1}{2}(\xi_{k+q} - \xi_k)(-J(q)+V(q))G(k+q\sigma,\tau=0) + J(-q)<\hat{S}(-q)\hat{S}(q)>J(q)$$

$$- 2\sigma V(-q)<\hat{\rho}(-q)\hat{S}(q)>J(q) + V(-q)<\hat{\rho}(-q)\hat{\rho}(q)>V(q) \quad (8)$$

The function $P(k,q,\sigma)$ will exhibit effects of correlations. Here $<\hat{S}(-q)\hat{S}(q)> \equiv <T_\tau \hat{S}(-q,\tau)\hat{S}(q,\tau-0^-)>$, $<\hat{\rho}(-q)\hat{\rho}(q)>$ and $<\hat{\rho}(-q)\hat{S}(q)>$ are similar to this expression. For simplification, we consider $T<T_c$ and $T \to T_c$, and get

$$F^+(k\sigma,\tau=0) = -\frac{1}{\beta}\sum_n [i\omega_n + \tilde{\xi}_{k\sigma} + \Sigma^{(+)}(k,i\omega_n)]^{-1} \cdot \frac{\Delta_+^{(-)}(k,i\omega_n)}{i\omega_n - \tilde{\xi}_{k\bar{\sigma}} - \Sigma^{(-)}(k,i\omega_n)}(1 - \frac{V(0)<\hat{\rho}(0)>}{-i\omega_n + \xi_k}) \quad (9)$$

where

$$\Sigma^{(\pm)}(k,i\omega_n) = \sum_q \frac{P(k,q,\sigma)}{i\omega_n \pm \xi_{k+q}}$$

$$\Delta^{(\pm)}(k,i\omega_n) = \frac{1}{2}\sum_q \frac{\xi_{k+q} - \xi_k}{-i\omega_n \pm \xi_{k+q}}[J(q)+V(q)]F(k+q\sigma,\tau=0) \quad (10)$$

$$\Delta_+^{(\pm)}(k,i\omega_n) = \frac{1}{2}\sum_q \frac{\xi_{k+q} - \xi_k}{-i\omega_n \pm \xi_{k+q}}[J(q)+V(q)]F^+(k+q\sigma,\tau=0)$$

To obtain an evident solution, we consider the in-site interaction $U$ is not too large. Because the function $F^+$ dominated by the frequency region where $\text{Im}\Sigma^{(+)}(k,\omega)=0$, Eq.(9) leads to

$$F^+(k\sigma,\tau=0) = \frac{1}{2}\frac{1}{E_{k\sigma,1}-E_{k\bar{\sigma},2}}[n_F(E_{k\sigma,1})\ z^{(+)}(E_{k\sigma,1})\ (1+\frac{V(0)<\hat{\rho}(0)>}{E_{k\sigma,1}-\xi_k})\sum_q \frac{\xi_{k+q}-\xi_k}{\xi_{k+q}+E_{k\sigma,1}}$$

$$-n_F(E_{k\bar{\sigma},2})\ z^{(-)}(E_{k\bar{\sigma},2})\ (1+\frac{V(0)<\hat{\rho}(0)>}{E_{k\bar{\sigma},2}-\xi_k})\sum_q \frac{\xi_{k+q}-\xi_k}{\xi_{k+q}+E_{k\bar{\sigma},2}}][J(q)+V(q)]F^+(k+q\sigma,\tau=0) \quad (12)$$

Where the spectral weight $z^{(\pm)}(\omega)=[1+\sum_q \frac{P(k,q,\sigma)}{(\omega\pm\xi_{k+q})^2}]^{-1}$. In the concrete, $\omega=E_{k\sigma,1}$ expresses the real solution of $\omega+\tilde{\xi}_{k\sigma}+\text{Re}\Sigma^{(+)}(k,\omega)=0$, and $\omega=E_{k\bar{\sigma},2}$ expresses the real solution of $\omega-\tilde{\xi}_{k\bar{\sigma}}-\text{Re}\Sigma^{(-)}(k,\omega)=0$, in which the spin index dependences are duo to the spin-charge correlation in Eq. (8).

Because $E_{k_F\sigma,1}\neq 0$ if $E_{k_F\bar{\sigma},2}\neq 0$, when the chemical potential was not within the energy bands $E_{k\bar{\sigma},1}$ and $E_{k\bar{\sigma},2}$, Eq. (12) will not give a large $T_c$, thus the large transition temperature requires $U$ is not too large. It is evident that there is the solution of $F^+(k_F)\neq 0$ for finite $T_c$ in Eq. (12). For example, when the chemical potential is located at the inside of excitation energies $E_{k\bar{\sigma},1}$ and $E_{k\bar{\sigma},2}$, Eq. (12) in Fermi surface gives this result

$$F^+(k_F,\tau=0) = \frac{1}{2k_BT_c}z^{(\pm)}(0)\frac{\xi_{k_F}-V(0)<\hat{\rho}(0)>}{\xi_{k_F}}\sum_q \frac{\xi_{k_F}-\xi_{k_F+q}}{\xi_{k_F+q}}[J(q)+V(q)]F^+(k_F+q,\tau=0) \quad (13)$$

Note $(\xi_{k_F}-\xi_{k_F+q})/\xi_{k_F+q}$ are either positive or negative for different $q$, we get $F^+(k_F)\neq 0$ for finite $T_c$. Because $z^{(\pm)}(0)$ decrease with $U$ while $J(q)+V(q)$ increase with $U$, the highest-$T_c$ will occur when $U$ is appropriate. Our results require $U$ is not excessively large, and they are against the Su's result [16]. Of course, Eq. (12) shows that the possible pairing is not limited at the Fermi surface, but superconductivity is dominated by the pairing at and around Fermi surface [22].

Having considered the depression of the possible ferromagnetism on the superconductivity, we conclude that the high-$T_c$ singlet superconductivity requires an antiferromagnetic exchange parameter $J_{ll'}$ and a positive parameter $V_{ll'}$ when the chemical potential is located within the energy-band of electron systems, the systems are in the metallic region.

Because $J$ and $V$ will directly contribute to spin-spin correlation and charge-charge correlation respectively, as shown in Eq. (8), we conclude that the spin-singlet superconductivity requires that both the spin correlation and the charge correlation are strong enough. The spin correlation and the charge correlation necessarily lead to the spin-charge correlation; therefore, the superconductivity requires that the spin-charge correlation is strong enough.

However, the condition displaying superconductivity also includes $F^+(k\sigma)=F^+(k\bar{\sigma})$ for the spin singlet pairing. When the spin correlation exists, the spin-charge correlation also exists, and then $P(k,q,\sigma)$ may depend on the spin index. Having substituted Eq. (8) into (10), we find $\Sigma(k,i\omega_n)=\Sigma_0(k,i\omega_n)+\sigma\Sigma_1(k,i\omega_n)$, this should lead the excitation energies to have such forms $E_\sigma(k)=E_0(k_x+\sigma Q_x, k_y+\sigma Q_y,...)+\sigma E_1(k)$. When the part $E_1(k)$ is large enough, the electron systems will show ferromagnetism, while this is impossible for our parameters in this article; when $Q_\alpha$ is large enough, the electron systems should show antiferromagnetism or spin density wave. If both $E_1(k)$ and $Q_\alpha$ are small or they reach some "matching" in quantity, the electron systems do not show any low-range magnetic order. Thus the long-range magnetic order could not exist when the spin-charge correlation is strong enough. On the contrary, magnetic orders suppress superconductivity. An interesting case is that $F^+(k\sigma)=F^+(k\bar{\sigma})$ may be met in higher temperature instead of lower temperature due to the possible non-monotonous temperature dependence of correlations. An example is ErRh$_4$B which undergoes the transition from superconductor to ferromagnetism with the decreased temperature [21], because the ferromagnetic correlation in this material may increase with the decreased temperature.

When the spin correlation is so large that $E_1(k)$ and $Q_\alpha$ can not counteract their effects on the pairing, superconductivity will disappear in this model due to $F^+(k\sigma)\neq F^+(k\bar{\sigma})$ at any finite temperature. This means that superconductivity requires some "balance" between spin correlation and charge correlation. That is why superconductivity usually occurs at the border of spin or charge orders. This result also leads us to conjecture that the competition between a spin order and a charge order must be strengthened when both the spin correlation and the charge correlation are strong. We can understand that an appropriate spin-charge correlation will lead high temperature superconductivity, because the spin-charge correlation could lead a strong spin fluctuation and a strong charge fluctuation, these strong fluctuations would induce tight-banded pairs which are responsible for the superconductivity. On the contrary, when $U$ is very small, the tight binding model is no longer in force, a popular basis set for these electron systems is plane waves, $J=0$ while $V_0(q)$ can be seen as the perturbed one, either spin correlation or charge correlation is weak as seen in electron gas, and Eq. (12) will give $T_c=0$. That is to say, superconductivity cannot occur in electron gas.

**In summary**, this work shows that the strong correlation favors the singlet superconductivity, and the high temperature superconductivity requires that the correlation is strong enough while the spectral weight around the chemical potential is also high enough. In other words, these calculations suggest that the high temperature superconductivity should be found in the metallic region near the Mott metal–insulator transition (MIT) if we use the notation from one band model. These works, and other authors such as in [24], also argue that spin orders (include SDW state) suppress superconductivity.

One may question these calculations; however, these conclusions on calculations are in agreement with the experiments.

Firstly, all high temperature superconductivities are in the metallic region near MIT. All copper-based p-type superconductors are from the so-called bandwidth-control insulators, and the optimal doped ones are in the metallic region near MIT. Some bandwidth-control MIT systems [25] do not show high temperature superconductivity, this is because the electron systems in these materials are far from the MIT, and they behave spin or charge orders.

Secondly, the strong correlation-dominated superconductivities originate from the electron-electron interaction renormalized by various factors, no matter what these superconductivities are mediated by ether spin or charge excitations. These have also explained why all high-temperature superconductivities could not be explained by a single kind of excitation, as questioned between physicists.

Thirdly, it seems all properties of high-temperature superconductors are consistent with this calculation. For example, we can conceive that the total correlation arrive at the strongest strength for the optimally doped p-type superconductors. Therefore, the little isotope effect is because the parameters of model are hardly changed by the isotope substitution, the T-linear resistivity is due to the dominated strong correlations[26], the considerable optical conductivity of low-frequency is because the chemical potential comes into the inside of the energy band, and the high-$T_c$ appear in these optimal cuprates as discussed above. Other factors, such as phonons and impurities, will play role in the properties of superconductors as soon as the total correlation is weak enough. For other behaviors of superconductors, such as the element substitution effect, the pressure effect, the pairing symmetry, they are consistent with this mechanism, although the detail discussion is not given in this work.

**ACKNOWLEDGMENTS**

We thank Nanjing University of Information Science & Technology for financial support.

State Communications149, (2009)261.